\documentstyle[preprint,aps]{revtex}
\tighten
\begin{document}
\draft
\title{Coulomb blockade at almost perfect transmission}
\author{K. A. Matveev}
\address{Massachusetts Institute of Technology, 12-105,
Cambridge, MA 02139}
\date{\today}
\maketitle
\begin{abstract}
  We study the equilibrium properties of a quantum dot connected to a bulk
  lead by a single-mode quantum point contact. The ground state energy and
  other thermodynamic characteristics of the grain show periodic
  dependence on the gate voltage (Coulomb blockade).  We consider the case
  of almost perfect transmission, and show that the oscillations exist as
  long as the transmission coefficient of the contact is less than unity.
  Near the points where the dot charge is half-integer the thermodynamic
  characteristics show a non-analytic behavior identical to that of the
  two-channel spin-$\frac12$ Kondo model. In particular, at any
  transmission coefficient the capacitance measured between the gate and
  the lead shows periodic logarithmic singularities as a function of the
  gate voltage.
\end{abstract}

\section{Introduction}
\label{sec:Introduction}

The phenomenon of Coulomb blockade of tunneling has recently attracted
a lot of interest, both theoretical and experimental\cite{reviews}.
It can be observed, e.g., by measuring conductance of a system of two
macroscopic leads connected to a small metallic grain by tunnel
junctions. At low temperature tunneling of an electron into the grain
leads to an increase of the electrostatic energy of the system by
finite amount $E_C=e^2/2C_0$, where $C_0$ is the grain capacitance.
Thus the tunneling conductance becomes exponentially small, with the
activation energy $E_C$. One can then add a gate electrode in order to
control the electrostatic energy
\begin{equation}
      E_Q = \frac{(Q - eN)^2}{2C_0}.
\label{electrostatics}
\end{equation}
Here $Q$ is the grain charge; parameter $N$ is proportional to the gate
voltage $V_g$. The activation energy is now a function of $V_g$. At the
values of the gate voltage corresponding to $N=n+\frac12$ the energies of
states with charges $en$ and $e(n+1)$ are equal, and the activation energy
vanishes. Therefore one observes periodic peaks of conductance as a
function of the gate voltage.

Recently the Coulomb blockade was observed in semiconductor
heterostructures\cite{Kastner}. Unlike in metallic systems, in a
semiconductor device it is often possible to control the barrier height by
adjusting voltage on additional gate electrodes. In such experiments one
can study the evolution of the Coulomb blockade as the transmission
coefficient ${\cal T}$ of the tunnel barrier changes from 0 to 1. The
experiments\cite{Vaart,Pasquier} indicate that the increase of the
transmission coefficient leads to the suppression of Coulomb blockade. At
${\cal T}\sim 1$ instead of well separated peaks weak periodic
oscillations of conductance $G(V_g)$ are observed. Experiment\cite{Vaart}
indicates that the Coulomb blockade disappears at ${\cal T} = 1$. On the
other hand, in the experiment\cite{Pasquier} the Coulomb blockade
oscillations were observed even when the conductance of the junction
exceeded $e^2/\pi\hbar$. To resolve this contradiction one needs a theory
of the Coulomb blockade in the regime of strong tunneling, ${\cal T}\to1$.

In this paper we study the Coulomb blockade in a quantum dot connected by
a controllable tunnel junction to a {\em single\/} electrode,
Fig.~\ref{fig:1}a.  Conductance measurements in such a system are not
possible. However, the Coulomb blockade shows up in the oscillations of
the equilibrium characteristics of the system, e.g., its ground state
energy $E$, or the average charge $\langle Q \rangle$ of the dot.
Experimentally the capacitance $C=\partial^2 E /\partial V_g^2$
between the gate and the lead can be measured\cite{Ashoori}.

Most of the theoretical work on Coulomb blockade is devoted to the case of
weak tunneling, when the transmission coefficient of the tunnel barrier is
small: ${\cal T}\ll 1$. At zero temperature, in the limit of very high
barrier the charge of the dot $Q(N)$ is quantized in units of the
elementary charge $e$, except for the degeneracy points $N=n+\frac12$,
where $Q$ changes from $ne$ to $(n+1)e$ (solid line in Fig.~\ref{fig:2}).
However, if the small probability of tunneling through the barrier is
taken into account, the charge of the dot is no longer a good quantum
number.  As a result the plateaus in $\langle Q(N)\rangle$ are not
horizontal, with the slope proportional to ${\cal T}$, Ref.\
\onlinecite{Glazman}. This phenomenon is due to the quantum fluctuations
of the dot charge caused by virtual processes of electron tunneling
between the grain and the lead.  Furthermore, the quantum fluctuations
were shown\cite{Matveev} to smear the steps of the average grain charge
$Q$ at half-integer values of $N$, making $\langle Q(N)\rangle$ a
continuous function (dashed line in Fig.~\ref{fig:2}).

In this paper we present a theory of the Coulomb blockade near the
strong tunneling limit ${\cal T} = 1$. In Sec.~\ref{sec:1D} we show
that the Coulomb blockade in the system shown in Fig.~\ref{fig:1}a is
described by a one-dimensional (1D) model. This allows us to use the
bosonization approach, and treat the Coulomb interaction exactly. In
agreement with experiment\cite{Vaart}, we find no contributions which
are periodic in $V_g$ in any measurable characteristic of the system
at ${\cal T}=1$. The backscattering on the barrier at ${\cal T}<1$ can
be treated in the bosonization approach as a small perturbation. In
Sec.~\ref{sec:Perturbation} we calculate the first non-vanishing
correction to the ground state energy of the system $E(N)$, average
grain charge $\langle Q(N)\rangle$, and capacitance $C(N)$. These
corrections are periodic in $N$, with the period corresponding to the
change of the grain charge by $e$. In the case of electrons with spin
the corrections diverge logarithmically at low energies, indicating
that the higher-order calculation is necessary. Such a calculation
performed in Sec.~\ref{sec:Exact} removes the singularities at all
values of $N$ except half-integer ones. In Sec.~\ref{sec:Kondo} we
discuss the non-analytic behavior of the thermodynamic characteristics
of the system at $N=n+\frac12$ using the analogy\cite{Matveev} between
the Coulomb blockade problem and the two-channel Kondo model. We argue
that the periodic logarithmic singularities in the capacitance
measured between the gate and the lead should be observed at {\em
any\/} value of the transmission coefficient ${\cal T}<1$.

\section{One-dimensional model}
\label{sec:1D}

The system we study is shown in Fig.~\ref{fig:1}a. The dot is connected to
the lead by a narrow constriction formed by applying voltage $V_a$ to the
auxiliary gates. We assume that the width of the constricton in its center
allows only a single transverse state below the Fermi level. In this sense
the electron gas inside the constriction is one-dimensional. As the
electron moves away from the center of the constriction, the the channel
becomes wider, Fig.~\ref{fig:1}b, and the number of transverse modes
grows.  Since the constriction is formed electrostatically, its boundaries
are smooth and do not scatter the electrons\cite{Lesovick}. Thus the
constriction creates an ideal quantum point contact between two
two-dimensional (2D) regions: the dot and the lead.

In the following we will neglect the fact that the dot size is finite,
i.e., the infinite system shown in Fig.~\ref{fig:1}b will be considered.
The difference between the two systems is that an electron entering a
finite dot will eventually return back to the lead through the
constriction. The time $\tau$ between these two events is determined by
the inverse width of the discrete energy levels in the grain. In the case
of an ideal single-mode junction the width is equal to the level spacing
$\varepsilon$ in the dot, and $\tau\sim\hbar/\varepsilon$.  An important
difference between non-interacting and interacting systems is that the
latter has another energy scale, $E_C$. In particular, the typical
frequency of charge fluctuations is $E_C/\hbar$ (see, e.g.,
Sec.~\ref{sec:Perturbation}). Since in a 2D dot $E_C\gg\varepsilon$, the
characteristic time $\hbar/E_C$ at which the Coulomb blockade develops is
much shorter than $\tau$. In the following we will consider the limit
$\tau\to\infty$ corresponding to the infinite system.

An important property of the infinite system shown in Fig.~\ref{fig:1}b is
that it is essentially one-dimensional. To see that, let us consider wave
function $\Psi(x,y)$ of an arbitrary state penetrating the constriction.
Near the center of the constriction the wave function is one-dimensional:
\begin{equation}
   \Psi_k(x,y) = \phi_0(y) e^{ikx},
\label{Psi}
\end{equation}
where $\phi_0(y)$ is the wave function of the ground state for the
transverse motion, wavevector $k$ is determined by the energy
corresponding to $\Psi(x,y)$. Since we consider an ideal contact
characterized by the quantized value of conductance $G=e^2/\pi\hbar$, the
wave function (\ref{Psi}) does not have a scattered component $\phi_0(y)
e^{-ikx}$. Outside the constriction the wave functions have a much more
complicated form. In particular, they may be strongly affected by the
disorder present in the 2D leads.  Nevertheless, one can label any wave
function by a single parameter $k$. Thus the Hamiltonian of the system of
electrons penetrating the constriction can be written as
\begin{equation}
  H_0 = \int\!\left(E_0+\frac{\hbar^2 k^2}{2m}\right)
        a_k^\dagger a_k^{} dk,
\label{H_0}
\end{equation}
where $a_k^\dagger$ is the creation operator for the electron in state
$\Psi_k(x,y)$, and $E_0$ is the energy of the transverse motion
corresponding to the wave function $\phi_0(y)$.

It is worth noting that Hamiltonian (\ref{H_0}) does not describe the
whole system of 2D electrons. For instance, the electrons with energies
below $E_0$ do not penetrate the constriction and are not included in
$H_0$. However, since we are interested in electron transport through the
constriction, the existence of electron states confined in one of the
electrodes, and therefore omitted in (\ref{H_0}), does not affect our
results.

The Hamiltonian (\ref{H_0}) has a one-dimensional form. However, unlike in
a usual 1D system, the density of states is determined by the 2D leads and
is therefore energy-independent. As a result the physical properties of
the system at energy scales of order of Fermi energy $E_F$ cannot be
described by the 1D model. On the other hand, the low-energy properties of
the system, such as conductance at low voltage and temperature $eV, T\ll
E_F$, or Coulomb blockade that develops at energy scale $E_C\ll E_F$, can
be described by the 1D model.

To consider the Coulomb blockade at ${\cal T}<1$ we have to add scattering
potential $V(x,y)$. We assume that this potential is localized inside the
constriction, where one can use the simple 1D form (\ref{Psi}) of the wave
functions. Then the Hamiltonian describing such scattering also takes a 1D
form:
\begin{equation}
   H' = \frac{1}{2\pi}
        \int\!\!\!\int V(k-k') a_k^\dagger a_{k'}^{}dk dk'.
\label{H'}
\end{equation}
Here the 1D scattering matrix element $V(q)$ is determined as
\begin{equation}
   V(q) = \int\!\!\!\int V(x,y) |\phi_0(y)|^2 e^{-iqx}dx dy.
\label{V(q)}
\end{equation}

To complete our 1D formulation of the Coulomb blockade problem, we must
show that the interaction Hamiltonian also has a 1D form. Assuming good
screening within the 2D dot, we will describe the Coulomb interaction by
the charging energy (\ref{electrostatics}), with $Q$ being the charge
inside the dot. In the absence of a tunnel barrier the boundary of the dot
is not well defined; we will assume it to be at the center of the
constriction\cite{note}. To find explicit form of the charge operator, we
note that there is an obvious relation between $Q$ and the current
operator:
\begin{equation}
  \dot{Q}\equiv -\frac{i}{\hbar} [Q,H_0] = J(0),
  \label{charge-current}
\end{equation}
where $J(0)$ is the operator of current at point $x=0$. The current
operator can be obtained by integrating the standard expression for the
current density over the transverse coordinate $y$. The expression for the
current density at $x=0$ is local, and we can use the 1D form (\ref{Psi})
of the wave functions inside the constriction. Then the current operator
takes the form
\begin{equation}
   J = \frac{e\hbar}{4\pi m}
        \int\!\!\!\int (k+k')a_k^\dagger a_{k'}^{}dk dk',
\label{J}
\end{equation}
where $m$ is the electron mass. Both $H_0$ and $J(0)$ have 1D forms in
terms of operators $a_k$. Hence the charge operator found form
Eq.~(\ref{charge-current}) is also essentially one-dimensional:
\begin{equation}
   Q = -i\frac{e}{2\pi}
        \int\!\!\!\int\frac{a_k^\dagger a_{k'}^{}}{k-k'} dk dk'.
\label{Q}
\end{equation}

Equations (\ref{H_0}), (\ref{H'}), (\ref{electrostatics}), and (\ref{Q})
present a complete 1D Hamiltonian of the system in $k$-representation. It
is more convenient to treat this Hamiltonian in coordinate representation,
which can be obtained by making Fourier transformation to the new 1D
fermion operators:
\begin{equation}\psi(x)=\frac{1}{\sqrt{2\pi}}\int a_k e^{ikx}dk.
  \label{Fourier}
\end{equation}
Unlike the initial 2D wave functions $\Psi(x,y)$, these new operators are
completely one-dimensional.

As we already mentioned, we are interested in low-energy properties of the
system. Thus we can linearize the spectrum of electrons in Eq.~(\ref{H_0})
near the two Fermi points, and write the fermion creation operators in
terms of left- and right-moving fermions:
\mbox{$\psi(x)=\psi_L(x)+\psi_R(x)$}.  The three parts of the Hamiltonian
then transform to
\begin{eqnarray}
H_0&=&\hbar v_F\int\Bigl[\psi_L^\dagger(x)(i\nabla-k_F)\psi_L^{}(x)
\nonumber\\
   & &\hspace{3.5em}-\psi_R^\dagger(x)(i\nabla+k_F)\psi_R^{}(x)\Bigr]dx,
  \label{H_0coord}\\
H'&=&\int V(x)\psi^\dagger(x)\psi(x)dx,
  \label{H'coord}\\
H_C&=&\frac{(Q - eN)^2}{2C_0}.
  \label{H_Ccoord}
\end{eqnarray}
Here $v_F$ is the Fermi velocity; the 1D scattering potential $V(x)$
is obtained from the real 2D potential $V(x,y)$ by averaging over the
electron density $|\phi_0(y)|^2$ in the transverse direction. The
charge operator (\ref{Q}) takes the form
\begin{equation}
  Q=\frac{e}{2}\int\Bigl[\psi_L^\dagger(x)\psi_L^{}(x)
                        +\psi_R^\dagger(x)\psi_R^{}(x)\Bigr]
    \,{\rm sgn}\,x\, dx.
  \label{Qcoord}
\end{equation}
As expected, the operator (\ref{Qcoord}) has the simple meaning of the
charge transferred from the region $x<0$ to the region $x>0$.

To summarize, we established that the Coulomb blockade in a dot connected
to a bulk electrode by means of a quantum point contact can be treated as
a 1D problem\cite{also}. This will greatly simplify the following
discussion, since we can now bosonize the Hamiltonian
(\ref{H_0coord})--(\ref{Qcoord}) and treat the quartic in fermion
operators interaction term (\ref{H_Ccoord}) exactly.

\section{Perturbation theory in reflection amplitude}
\label{sec:Perturbation}

\subsection{Bosonized Hamiltonian}
\label{sec:bosonization}

The bosonization technique\cite{Haldane} is applicable whenever the system
behavior is determined by the low-energy properties of the 1D electron
system.  As we already mentioned, the typical energy scale at which the
Coulomb blockade becomes important is $E_C$. This energy is much lower
than the Fermi energy, and the condition of applicability of the
bosonization approach is satisfied. At low energies the electron system
can be viewed as an elastic medium.  Therefore the bosonized Hamiltonian
can be written in terms of two variables: the displacement of the elastic
medium $u(x)$ and its momentum density $p(x)$. The Hamiltonian
(\ref{H_0coord}) of non-interacting electrons takes the form:
\begin{equation}
   H_0 = \int \!\left\{\frac{p^2(x)}{2mn_0}
                    + \frac{1}{2}mn_0v_F^2[\nabla u(x)]^2
              \right\}dx.
\label{H_0bosonized}
\end{equation}
Here $m$ is the electron mass, $n_0=mv_F/\pi\hbar$ is the electron
density. The two fields, displacement $u(x)$ and momentum density $p(x)$,
satisfy the standard commutation relation
\begin{equation}
[u(x), p(y)] = i\hbar\delta(x-y).
  \label{commutator}
\end{equation}

To bosonize the interaction Hamiltonian (\ref{H_Ccoord}) we should find
the expression for the charge transferred through the point $x=0$. This
can be done by substitution of the bosonized formula for the electron
density $\psi_L^\dagger\psi_L^{}+\psi_R^\dagger\psi_R^{}\to -n_0\nabla u$
into Eq.~(\ref{Qcoord}).  Alternatively one can just note the obvious
relation $Q=en_0u(0)$ between charge and displacement. In either case we
get
\begin{equation}
   H_C = E_C [n_0 u(0) - N]^2.
\label{H_C}
\end{equation}
It is important to emphasize that the interaction term (\ref{H_C}) is
quadratic in bosonic variables, and the Hamiltonian $H_0+H_C$ of the
system without scattering potential can be diagonalized exactly.

We are primarily interested in the periodic dependence of the ground
state energy on the gate voltage, i.e., on $N$. One can easily see
that after the transformation
\begin{equation}
   u(x) \to u(x) + N/n_0
\label{shift}
\end{equation}
the Hamiltonian $H_0+H_C$ does not depend on $N$. Hence we expect the
Coulomb blockade to be completely suppressed by charge fluctuations in the
absence of scattering potential, i.e., at ${\cal T} = 1$ (thin line in
Fig.~\ref{fig:2}). To find the Coulomb blockade oscillations at ${\cal T}
< 1$, one has to consider the effect of scattering on the potential of the
barrier\cite{others}.

In the bosonization approach only the low-energy properties of
the system are considered. The scattering on a localized potential is
therefore characterized by two constants: the amplitudes of forward
and backward scattering $V(0)$ and $V(2k_F)$ given by
Eq.~(\ref{V(q)}). In boson representation the scattering potential
has the form:
\begin{equation}
   H'=-V(0)n_0\nabla u(0)
      - \frac{V(2k_F)}{\pi\hbar v_F}D\cos[2\pi n_0 u(0)].
\label{scattering}
\end{equation}
where $D$ is the high energy cut-off (bandwidth). The first term in
Eq.~(\ref{scattering}) does not change under the transformation
(\ref{shift}) and does not lead to a dependence of the ground state
energy on $N$. On the contrary, the backscattering term becomes
periodically dependent on $N$, and therefore leads to periodic
dependence of all the thermodynamic properties on gate voltage.
Below we calculate the first non-vanishing periodic correction to the
ground state energy due to the backscattering.

\subsection{Perturbation theory for spinless electrons}
\label{sec:spinless}

To find the first-order correction to the ground state energy one has to
calculate the average of the cosine term in Eq.~(\ref{scattering}) over
the ground state of the quadratic Hamiltonian (\ref{H_0bosonized}),
(\ref{H_C}).  This can be performed along the same lines as in the
Debye-Waller theory:
\begin{equation}
   \delta E_{1} = -\frac{1}{\pi}|r|D \cos[2\pi n_0\langle u(0)\rangle]
                  e^{-2\pi^2 n_0^2
                  \langle\!\langle u^2(0)\rangle\!\rangle}.
\label{debye}
\end{equation}
Here we introduced the reflection amplitude $r$. (In the first order in
backscattering potential $r=V(2k_F)/i\hbar v_F$.) From Eq.~(\ref{H_C})
it is obvious that $\langle u(0)\rangle=N/n_0$. In a 1D elastic medium
the average quantum fluctuation of the displacement
$\langle\!\langle u^2\rangle\!\rangle=
\langle u^2\rangle -\langle u \rangle^2$ diverges logarithmically due
to the low-energy phonons. In our case the phonons with energies below
$E_C$ are pinned down by the interaction term (\ref{H_C}),
see Ref.~\onlinecite{Ruzin}. Therefore the fluctuation of $u$ is large,
but {\em finite}. To find $\langle\!\langle u^2\rangle\!\rangle$ we will
explicitly diagonalize the Hamiltonian $H_0+H_C$. This is achieved by
the transformation
\begin{eqnarray}
u(x)&=&\frac{N}{n_0}+
       \frac{1}{\sqrt{\pi}}\int_0^\infty u_k\cos(k|x|-\delta_k)dk,
\label{u-transformed}\\
p(x)&=&\frac{1}{\sqrt{\pi}}\int_0^\infty p_k\cos(k|x|-\delta_k)dk.
\label{p-transformed}
\end{eqnarray}
Here the phase shift $\delta_k$ is defined as
\begin{equation}
\delta_k=\arctan\left(\frac{E_C}{\pi \hbar v_F k}\right).
  \label{delta}
\end{equation}
In Eqs.~(\ref{u-transformed}) and (\ref{p-transformed}) we neglected the
odd modes proportional to $\sin kx$ because they do not contribute to
$u(0)$ and therefore are decoupled from both interaction $H_C$ and
scattering $H'$. The new fields $u_k$ and $p_k$ satisfy standard
commutation relations $[u_k, p_{k'}]=i\hbar\delta(k-k')$. In terms of
these new fields the Hamiltonian $H_0+H_C$ takes the form
\begin{equation}
H_0+H_C=\int_0^\infty\left(\frac{p_k^2}{2mn_0}
        + \frac{1}{2}mn_0(v_Fk)^2u_k^2\right)dk.
  \label{diagonalized}
\end{equation}
It follows from Eq.~(\ref{u-transformed}) that the contribution of the
low-frequency modes to the displacement $u(0)$ is suppressed by the factor
$\cos\delta_k\sim \hbar v_F k/E_C$. This gives rise to the low-frequency
cut-off in the logarithmic integral for $\langle\!\langle
u^2(0)\rangle\!\rangle$. A simple calculation then gives
\begin{equation}
   \langle\!\langle u^2(0)\rangle\!\rangle =
   \frac{1}{2\pi^2n_0^2}\ln\left(\frac{\pi D}{\gamma E_C}\right).
\label{fluctuation}
\end{equation}
Here $\gamma=e^{\bf C}$, with ${\bf C}\approx 0.5772$ being the
Euler's constant. We can now find the periodic correction to the
ground state energy using Eq.~(\ref{debye}),
\begin{equation}
   \delta E_{1} = -\frac{\gamma}{\pi^2}|r|E_C \cos2\pi N.
\label{deltaE_1}
\end{equation}
As expected the amplitude of oscillations of the ground state energy
becomes of the order of $E_C$ at $|r|\sim 1$.

The average charge $\langle Q \rangle$ in the dot can now be found using
(\ref{H_Ccoord}),
\begin{equation}
   \langle Q \rangle
   =eN - \frac{e}{2E_C}\frac{\partial\delta E_{1}}{\partial N}
   =eN - \frac{\gamma}{\pi}e|r|\sin 2\pi N.
\label{charge}
\end{equation}
The period of oscillations corresponds to the change of the average number
of particles in the dot by one. At weak reflection, $|r|\ll1$, the
amplitude of the oscillations of charge is small (dash-dotted line in
Fig.~\ref{fig:2}).

Finally, the periodic correction to the capacitance measured between the
gate and the lead can be found as $\delta C_1=\partial^2 \delta E_1/
\partial V_g^2$. It also exhibits periodic oscillations as a function of
the gate voltage.

In the above discussion we completely ignored the spins of electrons.
The spinless case can probably be realized in an experiment in a high
magnetic field. However in the absence of magnetic field one should
take the spins into account. We will now demonstrate that the spin degree
of freedom affects the above results dramatically.

\subsection{Perturbation theory for electrons with spin}
\label{sec:spinful}

To take into account electron spins we consider a model with two
channels corresponding to two spin directions:
\begin{eqnarray}
   H_0 &=& \!\int\!\! \left\{\!\frac{p_1^2 + p_2^2}{2mn_0}
            + \frac{mn_0v_F^2}{2}\!\left[(\nabla u_1)^2\!
                                    + (\nabla u_2)^2\right]
              \!\right\}\!dx,
\label{H_0spins}\\
   H_C &=& E_C \{n_0 [u_1(0) + u_2(0)] - N\}^2,
\label{H_Cspins}\\
   H'&=&-\frac{1}{\pi}|r|D
        \{\cos[2\pi n_0 u_1(0)]+\cos[2\pi n_0 u_2(0)]\}.
\label{H'spins}
\end{eqnarray}
The expressions for $H_0$ and $H'$ are obtained by a straightforward
generalization of Eqs.~(\ref{H_0bosonized}) and (\ref{scattering}) to the
two-channel case.  (In Eq.~(\ref{H'spins}) we neglected the forward
scattering terms.) The two channels are coupled through the interaction
term (\ref{H_Cspins}) which depends only on the sum $u_1(0) + u_2(0)$
representing the total charge brought into the dot.  Thus it is natural to
transform the Hamiltonian to charge and spin modes \mbox{$u_{c,s}= (u_1\pm
  u_2)/\sqrt{2}$} (and similarly for momentum densities $p_{c,s}$).  In
the new variables the Hamiltonian (\ref{H_0spins})--(\ref{H'spins}) takes
the form:
\begin{eqnarray}
   H_0 &=& \!\int\!\! \left\{\!\frac{p_c^2 + p_s^2}{2mn_0}
            + \frac{mn_0v_F^2}{2}\!\left[(\nabla u_c)^2\!
                                    + (\nabla u_s)^2\right]
              \!\right\}\!dx,
\label{H_0rotated}\\
   H_C &=& E_C \left[\sqrt{2}\,n_0 u_c(0) - N\right]^2,
\label{H_Crotated}\\
   H'&=&-\frac{2}{\pi}|r|D
        \cos\bigl[\sqrt{2}\,\pi n_0 u_c(0)\bigr]
        \cos\bigl[\sqrt{2}\,\pi n_0 u_s(0)\bigr].
\label{H'rotated}
\end{eqnarray}

One can easily calculate the first-order correction to the ground state
energy following the discussion of the spinless case. An important
difference between these two cases is that now we have two modes, and only
one of them is pinned down by Coulomb interaction (\ref{H_Crotated}).
Therefore the quantum fluctuation $\langle\!\langle u_s^2\rangle\!\rangle$
of the displacement in the spin channel diverges logarithmically.  This
leads to a strong suppression of the oscillations of the ground state
energy. The amplitude of these oscillations is $|r|\sqrt{E_C\varepsilon}$,
where $\varepsilon$ is the low energy cut-off of the order of the level
spacing in the dot. In the limit of large dot ($\varepsilon \to 0$) the
first-order correction to the ground state energy vanishes.  Thus in order
to obtain the Coulomb blockade oscillations one has to perform the
calculation up to the second order in barrier potential.

The second-order correction to the ground state energy can be
presented in the form
\begin{equation}
   \delta E_2 =  \frac{1}{\hbar}{\rm Im} \int_0^\infty \!
                  \langle H'(t)H'(0) \rangle\, dt.
\label{secondorder}
\end{equation}
{}From the explicit form (\ref{H'rotated}) of the perturbation $H'$ it
follows that the correlator $\langle H'(t)H'(0) \rangle$ factorizes into
charge and spin parts. The spin part is easily calculated:
\begin{equation}
\langle\cos[\sqrt{2}\,\pi n_0 u_s(0,t)]
\cos[\sqrt{2}\,\pi n_0 u_s(0,0)]\rangle = \frac{1}{2iDt}.
  \label{spin-part}
\end{equation}
The slow decay of the correlator at large $t$ is due to the low-frequency
modes. In the charge channel the low-frequency components of $u(0)$ are
suppressed.  As a result at $t\gg \hbar/E_C$ the charge part of the
correlator $\langle H'(t)H'(0) \rangle$ saturates,
\begin{eqnarray}
\langle\cos[\sqrt{2}\pi n_0 u_c(0,t)]
\cos[\sqrt{2}\pi n_0 u_c(0,0)]\rangle
\nonumber\\
= \frac{2\gamma E_C}{\pi D}
  \left(\cos^2\!\pi N-\frac{\pi^2\hbar^2}{4E_C^2t^2}\sin^2\!\pi N\right).
  \label{charge-part}
\end{eqnarray}
The substitution of Eqs.~(\ref{spin-part}) and (\ref{charge-part}) into
the expression for the second-order correction (\ref{secondorder}) gives
the integral which diverges logarithmically at large $t$.  The divergence
can be cut off at $t\sim\hbar/\varepsilon$ (or at $t\sim\hbar/k_BT$ if the
correction to the free energy at a finite temperature is being
calculated). The result has the form
\begin{equation}
   \delta E_2 = -\frac{4\gamma}{\pi^3} |r|^2 E_C
                 \ln \left(\frac{E_C}{\varepsilon}\right)
                 \cos^2\!\pi N.
\label{E_2}
\end{equation}
In the limit of large dot $E_C/\varepsilon\to \infty$, and the
second-order result diverges. This indicates that the terms of higher
orders in $|r|$ should be taken into account.

\section{Higher-order calculation of the ground state energy}
\label{sec:Exact}

To proceed with the higher-order calculation we will first simplify
our Hamiltonian (\ref{H_0rotated})--(\ref{H'rotated}). Since the
logarithmic divergence arises at small energy scales $E\ll E_C$, we do
not have to treat the charge fluctuations exactly. At such low
energies the charge fluctuations are suppressed by the interaction
term, and one can replace $\cos\bigl[\sqrt{2}\,\pi n_0 u_c(0)\bigr]$
in Eq.~(\ref{H'rotated}) by its value averaged over the unperturbed
ground state. After this simplification the charge-related part of the
Hamiltonian completely decouples and can be excluded.

Another simplification is made possible by the fact that the barrier
potential depends only on the spin mode displacement at $x=0$. Therefore
the odd elastic modes proportional to $\sin kx$ are not coupled with $H'$
and can be excluded. To this end we will change the variables:
\mbox{$u_{e,o}(x) = [u_s(x)\pm u_s(-x)]/\sqrt{2}$}, and similarly for the
momentum densities $p_{e,o}(x)$. Thus we arrive at a Hamiltonian
\begin{eqnarray}
   H_0 &=& \!\int_0^\infty\!\! \left\{\!\frac{p_e^2}{2mn_0}
            + \frac{mn_0v_F^2}{2}(\nabla u_e)^2
              \!\right\}\!dx,
\label{H_0new}\\
   H'&=&-\left(\frac{8\gamma E_C D}{\pi^3}\right)^{1/2}|r|
        \cos\pi N\cos[\pi n_0 u_e(0)].
\label{H}
\end{eqnarray}
The bandwidth $D$ here should be taken of the order of the charging energy
$E_C$, because only at such small energies the charge fluctuations can be
neglected.

The Hamiltonian (\ref{H_0new}), (\ref{H}) is very similar to the
Hamiltonian (\ref{H_0bosonized}), (\ref{scattering}) of non-interacting
electrons in the presence of scattering potential.  The important
difference is that the cosine in Eq.~(\ref{H}) has twice smaller argument
than the one in Eq.~(\ref{scattering}). The latter represents a product of
two fermion operators $\psi^\dagger(0)\psi(0)$. Similarly, Eq.~(\ref{H})
can be interpreted as a sum of two fermion operators:
$\psi^\dagger(0)+\psi(0)$. To support this observation we will re-write
the Hamiltonian in terms of a new bosonic field
\begin{equation}
\Phi(x)=-\pi n_0 u_e(|x|)
        + \frac{{\rm sgn}\, x}{\hbar n_0} \int_0^{|x|}\!p_e(x')dx'.
  \label{Phi}
\end{equation}
Unlike the old variables $u_e(x)$ and $p_e(x)$ defined at $x>0$, the new
field $\Phi(x)$ is defined on the whole $x$-axis and has the same
number of degrees of freedom. The commutation relations for the new
variables have the form: $[\Phi(x), \Phi(y)]=i\pi{\,\rm sgn}\,(x-y)$.

In terms of the field $\Phi$ the Hamiltonian (\ref{H_0new}), (\ref{H})
takes the form:
\begin{eqnarray}
   H_0 &=& \frac{\hbar v_F}{4\pi}\int_{-\infty}^\infty
            [\nabla \Phi(x)]^2dx,
\label{H_0chiral}\\
   H'&=&-\left(\frac{8\gamma E_C D}{\pi^3}\right)^{1/2}|r|
        \cos\pi N\cos\Phi(0).
  \label{Hchiral}
\end{eqnarray}

Expression (\ref{H_0chiral}) obviously coincides with the well-known form
of the bosonized Hamiltonian of a 1D gas of non-interacting right-moving
electrons (see, e.g., Ref.~\onlinecite{Haldane}). In this case the
operator $\sqrt{D/2\pi\hbar v_F}\, e^{i\Phi(x)}$ is identified as the
electron annihilation operator $\psi(x)$.  Consequently, the Hamiltonian
(\ref{H_0chiral}), (\ref{Hchiral}) can be de-bosonized to the following
form:
\begin{equation}
  H_f=\int_{-\infty}^{\infty}\left[\xi_k b_k^\dagger b_k^{} -
      \lambda(b_k^\dagger +b_k^{})\right]dk,
\label{debosonized}
\end{equation}
where $b_k^{}$ and $b_k^\dagger$ are the new fermion operators in
$k$-representation, electron energy $\xi_k=\hbar v_F k$ is measured from
the Fermi level, and $\lambda=\sqrt{2\gamma \hbar v_F
E_C/\pi^3}\,|r|\cos\pi N$.

The Hamiltonian (\ref{debosonized}) contains a term linear in fermion
operators $b_k^{}$ and $b_k^\dagger$. If it is treated perturbatively, one
obtains the result (\ref{E_2}). However, it is possible to take into
account all the higher terms of the perturbation theory. This can be done
by transforming the Hamiltonian (\ref{debosonized}) to a quadratic form.
One such transformation based on the Jordan-Wigner transformation to a
spin chain was suggested by Guinea\cite{Guinea}. We will use a simpler
transformation:
\begin{equation}
   b_k = \left(c + c^\dagger\right) c_k.
\label{transformation}
\end{equation}
One can easily check that if $c$ and $c_k$ are fermion operators, the
operators $b_k$ have correct anticommutation relations. After this
transformation we get a Hamiltonian quadratic in fermion
operators
\begin{equation}
  H_q=\int_{-\infty}^{\infty}\!\left\{\xi_k c_k^\dagger c_k^{} -
      \lambda\!\left[c_k^\dagger\left(c + c^\dagger\right)
              + \left(c + c^\dagger\right) c_k^{}\right]\!\right\}dk.
\label{transformed}
\end{equation}

The Hamiltonian $H_q$ is very similar to the Hamiltonian of a resonant
impurity at the Fermi level. Unlike the latter, $H_q$ does not conserve
the number of particles and should be diagonalized by a Bogoliubov
transformation. As a result we get the diagonal form
\begin{equation}
  H_q=E + \int_0^\infty \xi_k
             \left(C_k^\dagger C_k^{}
                   +\tilde C_k^\dagger \tilde C_k^{}\right)dk,
  \label{diagonal}
\end{equation}
where $E$ is the ground state energy of our Hamiltonian.
The two branches of the excitation spectrum correspond to some linear
combinations of particle and hole states. One of the braches is not
affected by coupling to the impurity level: $\tilde C_k^{}=(c_k^{} +
c_{-k}^\dagger)/\sqrt{2}$. The other branch has some admixture of
operators $c^\dagger$ and $c$,
\begin{eqnarray}
C^{}_k &=& \frac{\xi_k}{\sqrt{\xi_k^2 + \Gamma^2}}
         \frac{c_k^{} - c_{-k}^\dagger}{\sqrt{2}}
         - \sqrt{\frac{\hbar v_F\Gamma}{2\pi(\xi_k^2 + \Gamma^2)}}
           \left(c+c^\dagger\right)
\nonumber\\
       & & +\frac{\Gamma}{\pi\sqrt{\xi_k^2 + \Gamma^2}}
            \int_{-\infty}^{\infty}\frac{d\xi_{k'}}{\xi_k-\xi_{k'}}
            \frac{c_{k'}^{} - c_{-k'}^\dagger}{\sqrt{2}},
  \label{Bogoliubov}
\end{eqnarray}
where the principal value of the integral is assumed. The parameter
$\Gamma=4\pi\lambda^2/\hbar v_F$ has the meaning of the width of the
resonant level.

The correction $\delta E$ to the ground state energy of the Hamiltonian
$H_q$ can be found, e.g., by averaging Eq.~(\ref{diagonal}) over the
unperturbed ground state,
\begin{equation}
\delta E = - \int_0^\infty \xi_k
               \langle C_k^\dagger C_k^{}\rangle_0^{} dk.
  \label{average}
\end{equation}
The resulting integral over $k$ is logarithmically divergent at the upper
limit due to the second term in the right-hand side of
Eq.~(\ref{Bogoliubov}). However, as we already mentioned, the bandwidth in
our Hamiltonian should be $E_C$. Unlike in the case of the perturbation
theory, the logarithmic integral now has an intrinsic cut-off $\Gamma$ at
low energies. This low-energy cut-off is due to the higher order terms in
$\lambda$.  As a result the correction to the ground state energy is now
finite, $\delta E = -(\Gamma/2\pi)\ln(E_C/\Gamma)$. In our original
notations this result has the form:
\begin{equation}
   \delta E = -\frac{4\gamma}{\pi^3} |r|^2 E_C
                 \ln \left(\frac{1}{|r|^2\cos^2\pi N}\right)
                 \cos^2\pi N.
\label{exact}
\end{equation}

To summarize, the chain of transformations has lead us from bosonized
Hamiltonian (\ref{H_0spins})--(\ref{H'spins}) to simple form
(\ref{transformed}). The transformations are exact at low energies. They
uncover the low-energy cut-off $\Gamma$ for the logarithmic divergence of
the second-order perturbation theory.

It is important to note that $\Gamma$ vanishes at half-integer values of
$N$. At these points the logarithm in Eq.~(\ref{exact}) diverges. Due to
the pre-factor, correction $\delta E$ is not divergent, but still has a
non-analytic behavior at $N=n+\frac12$. The non-analiticity shows up in
the capacitance $C=\partial^2 E/\partial V_g^2$ measured between
the gate and the lead:
\begin{equation}
   \delta C(N) =  \frac{8\gamma}{\pi} |r|^2 \beta ^2 E_C
                 \ln\!\left(\frac{1}{|r|^2\cos^2\pi N}\right)\!
                 \cos 2\pi N.
\label{capacitance}
\end{equation}
Here $\beta$ is the parameter controlling the relation between $N$ and the
gate voltage, $N=\beta V_g$; its value is determined by the system
geometry. One can easily see that the capacitance is logarithmically
divergent at half-integer $N$. The nature of these singularities is
discussed in the next Section.

\section{Analogy to the two-channel Kondo model}
\label{sec:Kondo}

The logarithmic divergence of the capacitance (\ref{capacitance}) at
half-integer $N$ was found for the case of strong tunneling. It is
instructive to compare it with the similar divergence in
capacitance\cite{Matveev} in the weak tunneling case:
\begin{equation}
  \delta C(N) = -2\beta^2 E_C
                 \frac{1}{|t|}\exp\left(\frac{\pi}{4|t|}\right)
                 \ln\left(\frac{1}{|2N-1|}\right).
  \label{old}
\end{equation}
Here $t$ is the transmission amplitude; expression (\ref{old}) is written
for the vicinity of the point $N=\frac12$. Both expressions
(\ref{capacitance}) and (\ref{old}) predict a logarithmic singularity at
$N=\frac12$, with the factors in front of the logarithm being of the same
order at $|t|\sim|r|\sim1$. It is therefore natural to conjecture that the
logarithmic divergencies of capacitance exist not only in the limiting
cases of weak and strong tunneling, but at {\em any\/} value of the
transmission coefficient.

To support this idea we shall first outline the arguments leading to the
divergence (\ref{old}) of capacitance in the weak tunneling case. The
solution\cite{Matveev} was based on the mapping of the Coulomb blockade
problem onto an anisotropic multichannel Kondo model. At $N$ close to
$\frac12$ one can consider the perturbation theory in tunneling amplitude
and neglect all the terms involving virtual states with a charge different
from $0$ and $1$. This restriction on the possible charge states is due to
the large charging energy associated with all other states.  Thus all the
relevant terms of the perturbation theory are constructed in such a way
that first an electron tunnels through the barrier from left to right,
changing the dot charge from $0$ to $1$, then another electron tunnels
from right to left returning the dot to the state with $Q=0$, then one
more electron tunnels from left to right leading to $Q=1$, and so on.  One
can note that the same structure of the perturbation theory takes place
for the Kondo model with anisotropic coupling $J_\perp(\sigma^+ S^- +
\sigma^- S^+)$. Instead of the two types of electrons, left and right, we
now have two other types, spin-up and spin-down. Furthermore, each
electron scattering on the impurity flips its own spin, e.g., from up to
down, and the spin of the impurity, down to up. This means that the next
electron scattered on the impurity has to flip its spin from down to up,
then from up to down, etc.  This leads to the same structure of the
perturbation theory as in the Coulomb blockade problem.

A small deviation of the system from the point $N=\frac12$ gives rise to
the energy difference between the states with $Q=0$ and $Q=1$. It is
completely analogous to the effect of magnetic field $h$ in the Kondo
problem. Thus the capacitance of the system $C\sim \partial^2 E/\partial
N^2$ is analogous to the magnetic susceptibility of the Kondo impurity
$\chi_i=\partial^2 E/ \partial h^2$. The latter is inversely proportional
to the Kondo temperature, which leads to the exponentially large factor in
Eq.~(\ref{old}).

Finally, the presence of the real spin of electrons (conserved in the
tunneling process) should be interpreted as a ``color'' for the
electrons in the Kondo model. Thus the spin-$\frac12$ case maps onto
the {\em two-channel\/} Kondo model. The latter is
known\cite{bethe,Affleck,Emery,Sengupta} to exhibit some
non-Fermi-liquid properties. These include the logarithmic divergence
of the susceptibility at $h=0$, resulting in the logarithmic
singularity in Eq.~(\ref{old}).

The result (\ref{old}) for the capacitance in the weak coupling regime was
obtained using the exact solution of the Kondo model\cite{bethe}. Another
approach based on the renormalization group treatment\cite{Nozieres}
allows one to find some low-energy properties of the system at any
coupling strength. The main idea of this technique is that at low energies
the effective coupling constant in the Kondo model grows, and the system
approaches the strong coupling fixed point. This fixed point is stable,
and the low-energy properties of the system are determined by the leading
irrelevant perturbation. Therefore the low-energy behavior is universal,
i.e., independent of the initial conditions. The stable fixed point for
the multichannel Kondo problem was studied in detail by conformal field
theory methods\cite{Affleck}. In particular, the logarithmic behavior of
the susceptibility in the two-channel Kondo model was rederived.

To apply this method to the Coulomb blockade problem one should first find
the fixed points. There are two obvious fixed points: ${\cal T}=0$ and
${\cal T}=1$. As we already discussed, the weak tunneling fixed point
${\cal T}=0$ corresponds to the weak coupling fixed point in the Kondo
model. This fixed point is therefore unstable.  It is then natural to
assume that the strong tunneling fixed point ${\cal T}=1$ maps to the
strong coupling fixed point of the Kondo model.  To test this hypothesis
one should first show that the ${\cal T}=1$ fixed point is stable. This
can be done by calculating the scaling dimension of the perturbation $H'$
in the Hamiltonian (\ref{H_0rotated})--(\ref{H'rotated}). The correlator
$\langle H'(t)H'(0)\rangle$ can be found from Eqs.~(\ref{spin-part}) and
(\ref{charge-part}). At half-integer values of $N$ (corresponding to zero
magnetic field in the Kondo model) it decays as $1/t^3$. Thus the scaling
dimension of the perturbation $H'$ is $\frac32$, and the fixed point is
stable.

The stability of the ${\cal T} = 1$ fixed point suggests that at
half-integer $N$ the system with any value of transmission coefficient
${\cal T}$ approaches the strong tunneling limit.  We therefore can argue
that periodic logarithmic divergences of the capacitance
(\ref{capacitance}) and (\ref{old}) are a general property of the system
and should be observed at any value of the transmission coefficient. In
particular, the divergence of capacitance at weak tunneling (\ref{old})
can be interpreted as a consequence of the divergences (\ref{capacitance})
in the strong tunneling limit.

The last statement can be tested by calculation of the so-called Wilson
ratio in the strong tunneling limit. Indeed, since the weak-tunneling
properties of the Coulomb blockade problem were found by mapping to the
Kondo model, one expects the correction to the heat capacity to have a
non-analytic temperature dependence $\delta c_i\sim k_BT\ln(E_C/k_BT)$,
and its ratio to the correction to the susceptibility to be
universal\cite{Affleck,Sengupta}. If the low-energy behavior of the weak
tunneling model is controlled by the strong tunneling fixed point, the
same must be valid for the model (\ref{H_0rotated})--(\ref{H'rotated}). A
straightforward calculation of the correction to the heat capacity in the
second order in $H'$ gives
\begin{equation}
\delta c_i = \frac{\pi\gamma|r|^2}{E_C}k_BT\ln(E_C/k_BT).
  \label{heat}
\end{equation}
Taking into account the relation $h=2E_C(N-\frac12)$ between the effective
magnetic field $h$ in the Kondo model and the deviation of $N$ from a
half-integer value, we can find the susceptibility from Eq.~(\ref{E_2}) as
$\delta \chi_i=(2E_C)^{-2} \partial^2 \delta E_2/\partial N^2$. The result
is
\begin{equation}
\delta \chi_i = \frac{2\gamma|r|^2}{\pi E_C}\ln(E_C/k_BT).
  \label{chi}
\end{equation}
The ratio $k_BT\delta \chi_i/\delta c_i$ now takes the universal value
$2/\pi^2$, in agreement with the theory of the two-channel Kondo
model\cite{Affleck,Sengupta}.

Finally, it is instructive to discuss the relation between our
quadratic Hamiltonian (\ref{transformed}) describing the low-energy
properties of the Coulomb blockade problem and the similar Hamiltonian
for the Toulouse limit of the two-channel Kondo model. The latter was
obtained by Emery and Kivelson\cite{Emery} and has the following form:
\begin{eqnarray}
H & = & iv_F\int_{-\infty}^{\infty}
            \psi^\dagger(x)\frac{\partial\psi(x)}{\partial x} dx
\nonumber\\
     &&+\frac{J_x}{\sqrt{2\pi a}} [\psi^\dagger(0) + \psi(0)]
        [d^\dagger -d] + h (d^\dagger d -1/2).
\label{emery}
\end{eqnarray}
Unlike in the Hamiltonian (\ref{emery}), the second part of
Eq.~(\ref{transformed}) is proportional to the weak magnetic field:
$\lambda\propto h$. However, one can still bring the Hamiltonian
(\ref{emery}) to the form identical to (\ref{transformed}). A linear
transformation of fermion operators analogous to (\ref{Bogoliubov})
allows one to absorb the second term of Eq.~(\ref{emery}) into the
kinetic energy term. After such a transformation the third term of
Eq.~(\ref{emery}) takes the form, which is identical to the second
term of Eq.~(\ref{transformed}) at low momenta.  Therefore at low
energies the Hamiltonians (\ref{transformed}) and (\ref{emery}) are
equivalent. This proves our conjecture that the strong tunneling fixed
point is identical to the strong coupling fixed point of the
two-channel Kondo model.

\section{Conclusion}
\label{sec:Conclusion}

In this paper we have studied the Coulomb blockade in the limit of the
barrier transmission coefficient ${\cal T}$ close to unity. The Coulomb
blockade oscillations of the dot charge and capacitance persist as long as
${\cal T}<1$. As ${\cal T}$ approaches unity, the sharp peaks in the
system capacitance transform into a weak oscillation with periodic
logarithmic divergences at the points where the dot charge is
half-integer.

The analogy between the Coulomb blockade and Kondo problem discussed in
Sec.~\ref{sec:Kondo} allowed us to conclude that the logarithmic
divergences in capacitance should be observed at any value of ${\cal T}$.
Of course, the exact calculation of the capacitance in terms of the
transmission and reflection amplitudes is possible only in the limits
${\cal T}\ll1$ and $1-{\cal T}\ll1$.

\acknowledgements

The author is grateful to B.~L. Altshuler, A. Furusaki, L.~I. Glazman,
P.~A. Lee, and E. Wong for stimulating discussions. The work was sponsored
by Joint Services Electronics Program Contract DAAL03-92-C-0001.

\begin{figure}
\caption{(a) Schematic view of a quantum dot connected to a bulk 2D
electrode. The dot is formed by applying negative voltage to the gates
(shaded). Solid line shows the boundary of the 2D electron gas (2DEG).
Electrostatic conditions in the dot are controlled by the gate voltage
$V_g$. Voltage $V_a$ applied to the auxiliary gates controls the
transmission coefficient ${\cal T}$ through the constriction.
(b) Constriction between two 2D regions. Inside the constriction
the wave functions have 1D form (\protect\ref{Psi}).}
\label{fig:1}
\end{figure}

\begin{figure}
  \caption{The average charge $Q$ of the dot as a function of
    dimensionless gate voltage $N$ at different values of transmission
    coefficient: ${\cal T}=0$ (solid line), ${\cal T}\ll1$ (dashed line),
    $1-{\cal T}\ll1$ (dash-dotted line), and ${\cal T}=1$ (thin line).}
  \label{fig:2}
\end{figure}


\begin{references}

\bibitem{reviews} See, e.g., the review D. V. Averin and K. K. Likharev,
  in {\it Mesoscopic Phenomena in Solids}, edited by B. Altshuler et al.
  (Elsevier, Amsterdam, 1991), p. 173, and references therein.

\bibitem{Kastner} M. A. Kastner, Physics Today {\bf 46}, 24 (1993).

\bibitem{Vaart} N. C. van der Vaart, A. T. Johnson, L. P. Kouwenhoven, D.
  J. Maas, W. de Jong, M. P. de Ruyter van Steveninck, A. van der Enden,
  C. J. P. M. Harmans, and C. T. Foxon, Physica B 189, 99 (1993).

\bibitem{Pasquier} C. Pasquier, U. Meirav, F. I. B. Williams, D. C.
  Glattli, Y. Jin, and B. Etienne, Phys. Rev. Lett. {\bf 70}, 69 (1993).

\bibitem{Ashoori} Such an experiment in a slightly different geometry was
  performed by R. C. Ashoori, H. L. Stormer, J. S. Weiner, L. N. Pfeiffer,
  K. W. Baldwin, and K. W. West, Phys. Rev. Lett. {\bf 71}, 613 (1993).

\bibitem{Glazman} L. I. Glazman and K. A. Matveev, Zh. Eksp. Teor.  Fiz.
  {\bf 98}, 1834 (1990) [Sov. Phys. JETP {\bf 71}, 1031 (1990)].

\bibitem{Matveev} K. A. Matveev, Zh. Eksp. Teor. Fiz. {\bf 99}, 1598
  (1991) [Sov. Phys. JETP {\bf 72}, 892 (1991)].

\bibitem{Lesovick} L. I. Glazman, G. B. Lesovik, D. E. Khmel'nitskii, and
  R.~I. Shekhter, Pis'ma Zh. Eksp. Teor. Fiz. {\bf 48}, 218 (1988) [JETP
  Lett. {\bf 48}, 238 (1991)].

\bibitem{note} The uncertainty of the boundary position is of the order of
  the constriction length $L$. Our results are not affected by this
  uncertainty if $L\ll\hbar v_F/E_C$.

\bibitem{also} A similar model has been used in
  K. Flensberg, Phys. Rev. B {\bf 48}, 11156 (1993).


\bibitem{Haldane} F. D. M. Haldane, J. Phys. C {\bf 14}, 2585 (1981).

\bibitem{others} Similar approach was used by A. Furusaki and N.  Nagaosa,
  Phys. Rev. B {\bf 47}, 3827 (1993), and by K. Flensberg\cite{also}.

\bibitem{Ruzin} L. I. Glazman, I. M. Ruzin, and B. I. Shklovskii, Phys.
  Rev. B {\bf 45}, 8454 (1992).

\bibitem{Guinea} F. Guinea, Phys.  Rev. B {\bf 32}, 7518 (1985).

\bibitem{bethe} A. M. Tsvelick and P. B. Wiegmann, Z. Phys. B {\bf 54},
  201, (1984); N. Andrei and C. Destri, Phys. Rev. Lett. {\bf 52}, 364
  (1984);

\bibitem{Affleck} I. Affleck and A. W. W. Ludwig, Nucl. Phys. {\bf B360},
  641 (1991).

\bibitem{Emery} V. J. Emery and S. Kivelson, Phys. Rev. B {\bf 46}, 10812
  (1992).

\bibitem {Sengupta} A. M. Sengupta and A. Georges, preprint LPTENS
93/46 (1993).

\bibitem{Nozieres} P. Nozi\`eres, J. Low Temp. Phys. {\bf 17}, 31 (1974);
  P. Nozi\`eres and A. Blandin, J. Phys. (Paris) {\bf 41}, 193 (1980).

\end{references}
\end{document}